\documentclass{moriond}
\usepackage{graphicx}	

\def\be{\begin{equation}}
\def\ee{\end{equation}}
\def\bea{\begin{eqnarray}}
\def\eea{\end{eqnarray}}

\newcommand{\Photo}{}

\begin{document}
\vspace*{4cm}
\title{The dynamical origin of GW190521 in young massive star clusters}

\author{ Marco Dall'Amico }

\address{Physics and Astronomy Department Galileo Galilei, University of Padova, Vicolo dell’Osservatorio 3, I–35122, Padova, Ita}

\maketitle\abstracts{
GW190521 is the most massive binary black hole (BBH) merger observed to date. Due to its peculiar properties, the origin of this system is still a matter of debate: several hints may favor a dense stellar environment as a birthplace. Here, we investigate the possible formation of GW190521-like systems via three-body encounters in young massive star clusters (YSCs) by means of direct \textit{N}-body simulations.}

\section{Introduction}

Intermediate mass black holes (IMBHs) bridge the gap between stellar-mass black holes (BHs) and super-massive black holes in the range $100-10^5\,$M$_{\odot}$.  On May 21, 2019, the Ligo and Virgo facilities detected the collision of a binary black hole (BBH) massive enough to produce the first IMBH candidate ever detected with gravitational waves (Abbott et al.\cite{ab1}$^{,}\,$\cite{ab2}). This event, named GW190521, originated from the merger of two BHs with mass $m_{1}\simeq 85\,$M$_{\odot}$ and  $m_{2}\simeq 66\,$M$_{\odot}$, and  produced a $\sim 140\,$M$_{\odot}$ remnant that lies in the intermediate mass range. One of the distinctive features of this system is that the primary BH has a $99\%$ probability of lying inside the pair-instability (PI) mass gap ($\sim{}60-120\,$M$_{\odot}$, Abbott et al.\cite{ab1}$^{,}\,$\cite{ab2}), i.e. the mass range in which no BH is expected to be produced by the collapse of a single star, as a consequence of the unstable oxygen-silicon burning phase experienced by the progenitor. Moreover, the observed precessing spin parameter  $\chi_{\mathrm{p}}\sim{}0.68^{+0.25}_{-0.37}$, and an effective spin parameter  $\chi_{\mathrm{eff}}\sim{}0.08^{+0.27}_{-0.36}$ ($90\%$ credible interval) of this system favour a precessing binary model with in-plane spin components and high spin magnitudes. Finally, some authors claim that the GW190521 waveform might be compatible with an eccentric binary at the time of the merger (Gayathri et al.\cite{ga}, Romero-Shaw et al.\cite{ro}, Bustillo et al.\cite{bu}, Abbott et al.\cite{ab2}). All these properties might favor the formation of GW190521 via dynamical interactions in a dense stellar environment rather than via isolated binary evolution. 

\section{Three-body simulations in a nutshell}

We simulated $2\times10^{5}$ three-body encounters between a BBH and a single massive BH using the direct \textit{N}-body code {\sc arwv} (Chassonnery et al.\cite{ch}). {\sc arwv} implements a post-Newtonian treatment of the equation of motions up to the $2.5$ order, and a relativistic kick prescription for merger remnants. We set our three-body scattering experiments in the massive young star clusters (YSCs) of Di Carlo et al.\cite{ugo19}$^{,}\,$\cite{ugo20a}, and generate the masses and the semi-major axis of our systems from the BH population of Di Carlo et al.. In this way, our sample includes also BHs with mass inside and above the PI gap produced by stellar collisions. Moreover, we specifically require the intruder mass to be above the lower end of the PI mass gap. Finally, we sampled the dimensionless spin parameter of each BH from a Maxwell-Boltzmann distribution with $\sigma_\chi{}=0.1$ as already done in Bouffanais et al.\cite{y19}, while its orientation is isotropic. A complete description of these initial conditions is presented in Dall'Amico et al.\cite{me}.

\section{GW190521-like BBH mergers}
\begin{figure}
\centering
\begin{minipage}{0.49\linewidth}
\centerline{\includegraphics[width=0.95\linewidth]{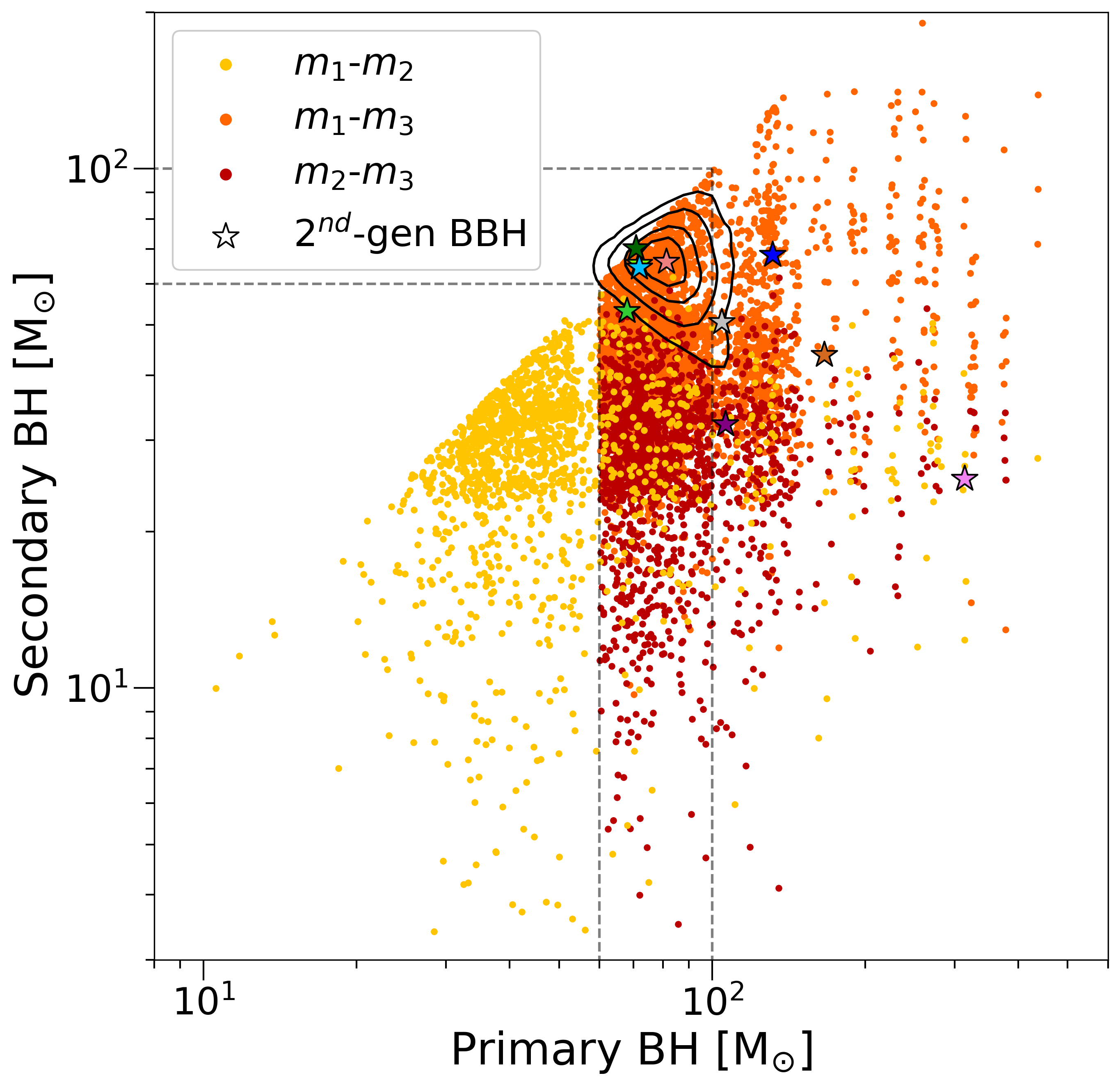}}
\end{minipage}
\begin{minipage}{0.49\linewidth}
\centerline{\includegraphics[width=0.95\linewidth]{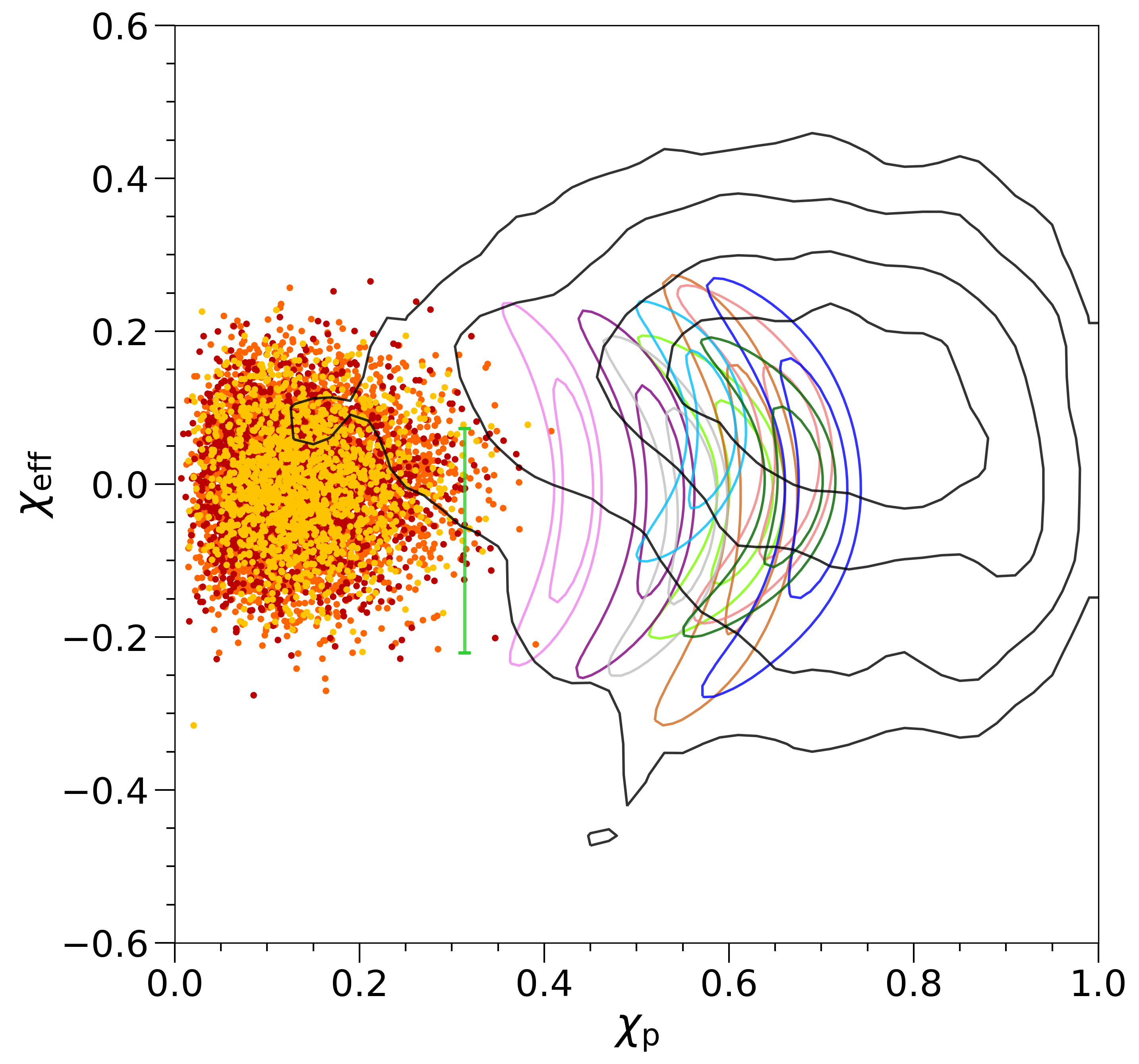}}
\end{minipage}
\caption[]{Left-hand panel: Primary and secondary masses of the simulated BBH mergers. Yellow circles are flyby BBHs, while orange (red) circles are exchanged BBHs where the intruder replaced the secondary (primary) BH. The black contour levels are the 25, 50, 75, 90$\%$ credible regions of GW190521 reported by Abbott et al.\cite{ab1}. Coloured stars are second-generation BBHs. The gray, salmon, cyan, lime-green and dark-green stars are inside the 90\% credible regions from Abbott et al.\cite{ab1}. The vertical dashed grey lines mark the lower-end of the PI mass gap, at $60$ M$_{\odot}$, and the lower end of the IMBH mass range, at $100$ M$_{\odot}$. Right-hand panel: Effective spin parameter $\chi_{\rm eff}$ versus precessing spin parameter $\chi_{\rm p}$ for all the BBH mergers. The colours are the same as the previous panel. The blue, salmon, dark-green, lime-green, cyan, purple, brown, grey and pink contours are the $50$ and $90\%$ credible regions for 9 out of the 10 second-generation BBHs. The green bar shows the last second-generation BBH for which $\chi_{\rm p}$ depends only on the spin of the first-generation component (see the main text for details). The black contours are the 25, 50, 75, 90$\%$ credible regions for the GW190521 spin parameters posterior reported by Abbott et al.\cite{ab1} and Abbott et al.\cite{ab2}.}
\label{fig:subfig}
\end{figure}

Our three-body simulations have three possible outcomes: flybys if the original binary ($m_1-m_2$) survived to the encounter, exchanges if the single BH replaced one component in the original binary ($m_1-m_3$ and $m_2-m_3$ if the secondary or the primary BH is kicked out) and ionizations. The left panel of figure \ref{fig:subfig} shows the primary and secondary BH masses of the BBH mergers at the end of the simulation as function of the outcomes. One every $\sim{}9$ BBH mergers ($11\%$ of the total) have both the primary and secondary mass inside the $90\%$ credible intervals of GW190521 ($85^{+21}_{-14}$ and $66^{+17}_{-18}$ M$_\odot$, Abbott et al.\cite{ab1}$^{,}\,$\cite{ab2}), the vast majority of which are exchanged BBHs ($98.5$\%). Most of these mergers are between $m_1$ and $m_3$ ($97.1$\%), while mergers between $m_2$ and $m_3$ are only the $1.4\%$ of the GW190521-like systems. Flybys and second-generation binaries  contribute to the $0.9\%$ and the $0.6\%$ of the GW190521-like systems. Specifically, five over all the ten second-generation BBHs lie inside the Abbott et al.\cite{ab1} $90\%$ credible regions for the  component masses of GW190521. In four of these simulations, the original binary experiences a strong encounter with the intruder BH, during which $m_{3}$ extracts enough internal energy from the binary to induce its merger. Despite the relativistic kick, the merger remnant resulting from this first coalescence forms a second-generation BBH with the intruder BH. These systems merge again in less than a Hubble time. The coalescence time of the inner binary $m_1-m_2$ computed at the beginning of the simulation is longer than the duration of the simulation (i.e., $10^{5}\,$yr) for all of these three mergers, meaning that the coalescence between $m_1$ and $m_2$ is sped up by the three-body interaction. Finally, one out of five second-generation BBHs matching the component masses of GW190521 is instead the product of an exchange event.

The right-hand panel of figure \ref{fig:subfig} shows the effective spin parameter $\chi_{\mathrm{eff}}$ as function of the precessing spin parameter $\chi_{\mathrm{p}}$ for all the BBH mergers. These quantities are computed with the following expressions:
\begin{eqnarray}\label{eq:xeff}
\chi_{\rm eff}= \frac{(m_{i}\,{}\vec{\chi}_{i}+m_{j}\,{}\vec{\chi}_{j})}{m_{i}+m_{j}}\cdot{}\frac{\vec{L}}{L}, \;\;\;\;\;\;\;\;
\chi_{\rm p}=\frac{c}{B_{i}\,{}G\,{}m_{i}^2}\,{}\max{(B_{i}\,{}S_{i\perp{}},\,{}B_{j}\,{}S_{j\,{}\perp})},
\end{eqnarray}
where $G$ is the gravity constant, $c$ the speed of light, $\vec{L}$ is the orbital angular momentum vector of the system, $S_{i\perp{}}$ and $S_{j\perp{}}$ are the spin angular momentum components in the orbital plane of the primary and secondary bodies of the binary, $B_i\equiv{}2+3\,{}q/2$ and $B_j\equiv{}2+3/(2\,{}q)$ with $q=m_{j}/m_{i}$ ($m_{i}\geq{}m_{j}$). Since dynamics randomly re-distributes the initial BH spins' orientation during a three-body interaction, we compute the final spin parameters $\chi_{\rm p}-\chi_{\rm eff}$ re-drawing the direction of each BH spin isotropically over a sphere but conserving their initial magnitude. For the BH remnants that pair up in second-generation BBHs we do not derive a single value but rather generate a full set of direction angles still sampled from an isotropic distribution. This implies that second-generation BBHs are represented in the plot as contour regions, with the exception of one system (green bar) in which the first-generation component has a higher spin magnitude than the second-generation companion, and thus dominates the $\chi_{\rm p}$ term in equation~\ref{eq:xeff} resulting in one single $\chi_{\rm p}$ value for a set of $\chi_{\rm eff}$ values. The plot highlights two distinct populations of mergers. First-generation BBHs, which underwent exchanges and flybys, cover the parameter space at low values of the precessing spin, while second-generation BBHs are located at high $\chi_{\rm p}$.  Half of all second-generation BBH mergers (five out of ten BBHs) match both the component masses and the spin parameters of GW190521 inside the 90\% credible intervals reported by Abbott et al.\cite{ab1}, while only $0.1\%$ of the first-generation BBH mergers  have both component masses and spin parameters inside the $90\%$ credible intervals of GW190521 according to Abbott et al.\cite{ab1}. This is an effect of our assumption that all first-generation BH's spin magnitudes are distributed according to a Maxwellian distribution with $\sigma{}_\chi{}=0.1$. Had we assumed a larger value for $\sigma{}_{\chi}$, we would have obtained a correspondingly higher fraction of first-generation BBHs matching GW190521's component masses and spin parameters.

The intersection of the two BBH samples that lie inside the posterior regions for the component masses (left panel, Figure~\ref{fig:subfig}) and spin parameters (right panel, Figure~\ref{fig:subfig}) of GW190521 contains twelve systems. These are five second-generation BBHs (marked by the gray, salmon, cyan, lime-green and dark-green systems in the plots) and seven exchanged binaries where $m_{3}$ replaced $m_{2}$ in the original system. The merger product of all these systems is an IMBH with a mass and a dimensionless spin magnitude inside the $90\%$ credible intervals of GW190521 ($M_{\rm rem}=142^{+28}_{-16}$ M$_\odot$ and $\chi_{\rm rem}=0.72^{+0.09}_{-0.12}$, Abbott et al.\cite{ab1}$^{,}\,$\cite{ab2}). Finally, two of the five second-generation BBHs that match the properties of GW190521 have respectively $e\sim0.003$ ($e\sim0.4$) and $e\sim0.004$ ($e\sim0.3$) at $10$ Hz ($10^{-2}$ Hz).

\section{Merger rate density}

We compute the approximate merger rate density of GW190521-like systems from our simulations as
\begin{eqnarray}\label{eq:eq11}
    \mathcal{R}_{\rm GW190521}\sim{0.03}\,{\rm Gpc}^{-3}\,{\rm yr}^{-1}\left(\frac{N_{\rm 190521}}{6}\right)\,{}\left(\frac{N_{\rm BBH}}{3606}\right)^{-1}\nonumber\,{}\left(\frac{\mathcal{R}_{\rm BBH}(z=0.8)}{170\,{\rm Gpc}^{-3}\,{}{\rm yr}^{-1}}\right)\,\left(\frac{f_{\rm YSC}}{0.7}\right)\,\left(\frac{f_{\rm corr}}{0.14}\right),
\end{eqnarray}
 where  $N_{\rm 190521}$ is the number of simulated BBH mergers with the mass of the components and the effective and precessing spin parameters inside the $90\%$ credible intervals reported by Abbott et al.\cite{ab1} and Abbott et al.\cite{ab2}, $N_{\rm BBH}$ is the number of BBH mergers in our simulations, $\mathcal{R}_{\rm BBH}(z=0.8)$ is the BBH merger rate density at $z\simeq{}0.8$ (i.e., the median redshift value of GW190521; Abbott et al.\cite{ab1}$^{,}\,$\cite{ab2}). We calculated $\mathcal{R}_{\rm BBH}$ for the YSCs simulated by Di Carlo et al.\cite{ugo20b} following the method described in Santoliquido et al.\cite{sa}. Finally, $f_{\rm YSC}$ is the fraction of BBH mergers that originate in YSCs, according to the fiducial model of Bouffanais et al.\cite{y21}, and $f_{\rm corr}$ is a correction factor to compensate for the bias we introduced when we simulated only intruders with $m_3\geq{}60$ M$_\odot$. The number of BBH mergers matching the effective and precessing spin parameters of GW190521 is strongly affected by our choice of the spin magnitude of first-generation BHs, which is drawn from a Maxwellian distribution with $\sigma_{\chi}=0.1$. A choice of $\sigma_{\chi}=0.2$ would have produced $109$ first-generation BBH mergers with the same properties as GW190521, rather than just three binaries as derived with $\sigma_{\chi}=0.1$. Hence, the merger rate density of GW190521-like systems  is very sensitive to the spin distribution of first-generation BBHs: we obtain  $\mathcal{R}_{\rm GW190521}\sim{0.01}\,$Gpc$^{-3}\,$yr$^{-1}$ if $\sigma_\chi=0.01$ (no first-generation BBH mergers matching GW190521) and $\mathcal{R}_{\rm GW190521}\sim{0.6}\,$Gpc$^{-3}\,$yr$^{-1}$ if $\sigma_\chi=0.2$.

\section*{Summary}
In this work we report the results of the three-body simulations performed by Dall'Amico et al.\cite{me}. Out of a total of $10^5$ simulated binary-single encounters, only in 12 cases the interaction produced a BBH merger that match the main properties of GW190521 ($m_1$, $m_2$, $\chi_{\rm eff}$, $\chi_{\rm p}$, M$_{\rm rem}$, $\chi_{\rm rem}$) within the $90\%$ credible interval reported by Abbott et al.\cite{ab1}. Our results imply that, if GW190521 was born in a massive YSC, it is either a BBH resulting from an exchange with a massive intruder ($\geq60\,$M$_{\odot}$) or a second-generation BBH merger triggered by a resonant three-body encounter. We find a merger rate density of $\mathcal{R}_{\rm GW190521}\sim{0.03}\,$Gpc$^{-3}\,$yr$^{-1}$ for GW190521-like BBHs formed via three-body encounters in YSCs. This value lies within the $90\%$ credible interval derived by Abbott et al.\cite{ab1}, but it strongly depends on the prescription adopted for the initial BH spin distribution. 


\section*{Acknowledgments}

MD acknowledges financial support from Cariparo foundation under grant 55440. This work is partially supported by the European Research Council for the ERC Consolidator grant DEMOBLACK, under contract no. 770017.

\end{document}